\documentclass[twocolumn,trackchanges]{aastex631}

\graphicspath{{./}{figures/}}

\begin{document}

\title{Atmospheric retrievals suggest the presence of a secondary atmosphere and possible sulfur species on L\,98-59\,d from JWST NIRSpec G395H transmission spectroscopy}

\author[0000-0002-9124-6537]{Agnibha Banerjee}\thanks{E-mail: agnibha.banerjee@open.ac.uk}
\affiliation{School of Physical Sciences, The Open University, Milton Keynes, MK7 6AA, UK}

\author[0000-0003-3726-5419]{Joanna K. Barstow}
\affiliation{School of Physical Sciences, The Open University, Milton Keynes, MK7 6AA, UK}

\author[0000-0003-0854-3002]{Amélie Gressier}
\affiliation{Space Telescope Science Institute, 3700 San Martin Drive, Baltimore, MD 21218, USA}

\author[0000-0001-9513-1449]{Néstor Espinoza}
\affiliation{Space Telescope Science Institute, 3700 San Martin Drive, Baltimore, MD 21218, USA}

\author[0000-0001-6050-7645]{David K. Sing}
\affiliation{William H. Miller III Department of Physics and Astronomy, Johns Hopkins University, Baltimore, MD 21218, USA}

\author[0000-0002-0832-710X]{Natalie H. Allen}
\affiliation{William H. Miller III Department of Physics and Astronomy, Johns Hopkins University, Baltimore, MD 21218, USA}

\author[0000-0001-7058-1726]{Stephan M. Birkmann}
\affiliation{European Space Agency, European Space Astronomy Centre, Camino Bajo del Castillo s/n, E-28692 Villanueva de la Cañada, Madrid, Spain}

\author[0000-0002-8211-6538]{Ryan C. Challener}
\affiliation{Department of Astronomy, Cornell University, 122 Sciences Drive, Ithaca, NY 14853, USA}

\author[0000-0001-7866-8738]{Nicolas Crouzet}
\affiliation{Leiden Observatory, Leiden University, P.O. Box 9513, 2300 RA Leiden, The Netherlands}

\author[0000-0002-8050-1897]{Carole A. Haswell}
\affiliation{School of Physical Sciences, The Open University, Milton Keynes, MK7 6AA, UK}

\author[0000-0002-8507-1304]{Nikole K. Lewis}
\affiliation{Department of Astronomy and Carl Sagan Institute, Cornell University, 122 Sciences Drive, Ithaca, NY 14853, USA}

\author[0000-0001-7237-6494]{Stephen R. Lewis}
\affiliation{School of Physical Sciences, The Open University, Milton Keynes, MK7 6AA, UK}

\author[0009-0006-2395-6197]{Jingxuan Yang}
\affiliation{Atmospheric, Oceanic and Planetary Physics, Department of Physics, University of Oxford, Oxford OX1 3PU, UK}
\begin{abstract}

L\,98-59\,d is a Super-Earth planet orbiting an M-type star. We performed retrievals on the transmission spectrum of L\,98-59\,d obtained using NIRSpec G395H during a single transit, from JWST Cycle 1 GTO 1224. The wavelength range of this spectrum allows us to detect the presence of several atmospheric species. We found that the spectrum is consistent with a high mean molecular weight atmosphere. The atmospheric spectrum indicates the possible presence of the sulfur-bearing species H$_2$S and SO$_2$, which could hint at active volcanism on this planet if verified by future observations. We also tested for signs of stellar contamination in the spectrum, and found signs of unocculted faculae on the star. The tentative signs of an atmosphere on L\,98-59\,d presented in this work from just one transit bodes well for possible molecular detections in the future, particularly as it is one of the best targets among small exoplanets for atmospheric characterization using JWST.

\end{abstract}


\section{Introduction} \label{sec:intro}

The era of detection and characterization of atmospheres
around rocky exoplanets is now here. \citet{LustigYaeger23, Lim2023, Moran2023, May2023, Kirk2024, damianoLHS1140Potentially2024} have already used the immense capabilities of JWST to
measure the transmission spectra of the atmospheres of rocky
exoplanets.
However, statistically significant detectable absorption
features have been scarce - owing to the low atmospheric
scale heights of rocky planets, possible cloud cover,
and stellar contamination from the M-dwarf host stars \citep{Moran2023}.

Here, we conducted atmospheric retrievals on the transmission
spectrum of L\,98-59\,d \citep{l9859_kostov2019, Cloutier2019}. The L\,98-59 system consists of
two more rocky transiting planets and another non-transiting planet \citep{Demangeon2021}.
Previous attempts to measure the transmission spectra between 1.1 -- 1.7 $\mu$m of planet b \citep{Damiano2022}, planet c \citep{barclayTransmissionSpectrumPotentially2023}, and planet d \citep{zhouHubbleWFC3Spectroscopy2023} using HST WFC3 have 
not ruled out the possibility of high mean molecular weight or cloudy atmospheres.
\citet{zhouHubbleWFC3Spectroscopy2023} ruled out cloud-free Hydrogen and Helium atmospheres for planets c and d. However, they could not exclude the possibility of primary cloudy/hazy or water-rich atmospheres.

L\,98-59\,d is an ideal
target for atmospheric studies due to its high Transmission Spectroscopy
Metric \citep{tsm} value. It may also sustain volcanism on its surface \citep{seligman24}, providing a viable source for a secondary atmosphere or a mixed primary and secondary, or hybrid \citep{tianAtmosphericChemistrySecondary2024} atmosphere.

L\,98-59\,d, with a radius of 1.52 $R_{\oplus}$, a mass of 1.94 $M_{\oplus}$, and an equilibrium temperature of 416 K is inconsistent with a pure rocky (Earth-like) composition and is potentially too small to be explained by H-He gas accretion from the protoplanetary disk \citep{Luque2022}. All three planets in the L\,98-59 system receive substantial X-ray and extreme-ultraviolet flux, leading to rapid water loss that significantly affects their developing climates and atmospheres \citep{Fromont2024}. While there is no solid claim of an atmosphere on L\,98-59\,c, planet d is closer to the cosmic shoreline and more likely to possess an atmosphere. L\,98-59\,d's density categorizes it as a `Water-world' according to \citet{Luque2022}, suggesting a composition of rock and water ice in roughly equal proportions by mass. \citet{seligmanPotentialMeltingExtrasolar2024} state that it may be possible to detect at atmosphere on L\,98-59\,d with 3-5 transits using NIRSpec/G395H, that could contain some amount of SO$_2$.

We analyzed a 2.9 - 5.2 $\mu$m JWST transmission spectrum obtained 
using NIRSpec \citep{Birkmann2022, Ferruit2022} G395H. We performed several sets of atmospheric retrievals with varying assumptions on
this spectrum to infer the atmospheric composition of L\,98-59\,d.
An atmospheric retrieval solves the inverse problem of going from
a measured spectrum to the range of atmospheric properties consistent with the observations. This is done by 
comparing the observed spectrum with large numbers of generated 
spectra  to explore the parameter space and find the best fitting combinations of model atmosphere properties \citep{Barstow2017, Tsiaras2018, grantJWSTTSTDREAMSQuartz2023,beattySulfurDioxideOther2024, bennekeJWSTRevealsCH2024, holmbergPossibleHyceanConditions2024, huSecondaryAtmosphereRocky2024}.

We used the transmission spectrum obtained for the JWST GTO
1224 program (PI: Birkmann) using NIRSpec G395H for 1 transit of L\,98-59\,d on 25
June 2023. The observations were reduced using two different pipelines
- \texttt{transitspectroscopy} \citep{Espinoza2022} and \texttt{FIREFLy}
\citep{Rustamkulov2022, Rustamkulov2023}, to obtain transmission spectra by binning the transit depths to a resolution of R$\sim$100. The reduction
methodology is discussed in detail in an accompanying paper from the GTO 1224 collaboration \citep{Gressier2024}, hereafter Paper I.
The \texttt{transitspectroscopy} reduction is used for
the majority of the retrievals in this letter. For completeness, we also ran a retrieval on the \texttt{FIREFly} reduction to explore the influence of data reduction on the retrieved parameters.

In Section \ref{sec:methods} we describe the retrieval setup and the priors used for each parameter.
In Section \ref{sec:Atm_retr} we define the atmospheric models used for retrievals and discuss the results from the suite of
retrievals that we have performed. In Section \ref{sec:discuss} we consider the significance of the presence of sulfur species in this atmosphere, compare results from the two independent data reductions presented in Paper I, and discuss the potential for future observations.

\section{Methods} \label{sec:methods}
\subsection{Retrieval Setup with \texttt{NEMESISPY}}
NEMESIS (Non-linear optimal Estimator for Multivariate spectral analySIS)
\citep{Irwin2008} is a versatile retrieval tool widely employed in the 
study of planetary atmospheres, both within our Solar System and beyond \citep{Barstow2016a, irwinDetectionHydrogenSulfide2018,braude2020, Barstow2020, irwin5D2020, irwinLatitudinalVariationMethane2021,singWarmNeptuneMethane2024}.
It utilizes a fast correlated-k method to solve the radiative transfer 
equation involving multiple absorbing gases. In this work, we used the
Python adaptation of the Fortran-based NEMESIS, \texttt{NEMESISPY} \citep{Yang23, yangSimultaneousRetrievalOrbital2024, yangNEMESISPYPythonPackage2024}.

We employed the PyMultiNest nested sampling solver \citep{Buchner2014,
feroz1, Feroz2009, Feroz2019} to explore potential solutions. 
Nested Sampling \citep{Skilling2004, skilling_nested_2006} is a
computational method for estimating the marginal likelihood of a
model and performing Bayesian parameter estimation. It sequentially
samples the prior distribution by generating live points, and
gradually fills the parameter space with points of increasing 
likelihood. In our study, we configured it with 2000 live points
and an evidence tolerance value of 0.5.

For the gas opacities, we used k tables (R=1000) from ExoMol \citep{exomol, exomol_katy} 
encompassing the following molecular species: H$_2$O \citep{polyansky2018exomol},
CO$_2$ \citep{Yurchenko2020}, CO \citep{li2015rovibrational}, NH$_3$ 
\citep{ExoMol_NH3}, PH$_3$ \citep{sousasilva2014}, CH$_4$ \citep{ExoMol_CH4},
SO$_2$ \citep{ExoMol_SO2}, and H$_2$S \citep{ExoMol_H2S}. These k tables 
are then channel-averaged to match the resolution of the data prior
to the retrieval process \citep{Irwin2020}. We evaluate Rayleigh scattering cross sections using data from \citet{Allen1976}. Additionally, we incorporate collision-induced
continuum absorption arising from H$_2$-H$_2$
\citep{borysow89, borysowfm89, 
Borysow2001, borysow02, Rothman2013},
and N$_2$-N$_2$ \citep{lafferty1996}.

\subsection{Atmospheric Model}\label{sec:priors}
In our setup, the atmosphere was divided into 100 equal intervals in log-pressure, starting from $10^{1}$ atm and extending to $10^{-7}$ atm. To remain agnostic of the atmosphere's
background composition, we used Centered Log-Ratio (CLR) \citep{Benneke2012}
priors for the gas abundances. The CLR parameterization treats all chemical species in the model equally, enabling any of them to be the dominant atmospheric species. This allows for a flexible exploration of possible background compositions. The CLR parameterization is a new addition to \texttt{NEMESISPY}, implemented in a manner similar to \texttt{POSEIDON} \citep{MacDonald2017, MacDonald2023}. Our implementation of CLR priors is non-uniform in the CLR space, and thus avoids biases in results as described in \citet{Damiano2021}. In the log volume mixing ratio or log(VMR) space, this manifests as a distribution peaked in the high values, and a flat tail towards the low values. We use $10^{-12}$ as the lower limit for the volume mixing ratios.


Clouds were represented using an opaque cloud top ($P_{\text{top}}$), with the prior for its logarithm, log($P_{\text{top}}$) defined as $\mathcal{U}(-7, 1)$ atm,
and power-law scattering due to hazes above the cloud layer \citep{MacDonald2017}, with the prior for power defined as $\mathcal{U}(-8, 4)$, where $\mathcal{U}(a, b)$ is a uniform distribution with a and b as the upper and lower bounds. We also include a Rayleigh enhancement parameter named Hazemult, with priors defined as $\mathcal{U}(-5, 5)$.
We used an isothermal Temperature-Pressure Profile, unless otherwise specified. The prior for the isothermal
temperature was defined as $\mathcal{U}(311, 625)$ K. This choice is further explained in Section \ref{temp_constrain}.
The ratio of the radius of the planet at 10 atm to the white-light 
transit radius of the planet was left as a free parameter ($f_{\text{ref}}$) with the prior set as $\mathcal{U}(0.7, 1.3)$. Following \citet{aldersonEarlyReleaseScience2023} and \citet{May2023}, we allowed for an offset between the NRS1 and NRS2 wavelength ranges. The prior distribution for this offset in ppm is set as $\mathcal{N}(0, 40)$, where $\mathcal{N}(\mu, \sigma)$ is a normal distribution with a mean of $\mu$ and standard deviation of $\sigma$. The values of stellar effective temperature, stellar radius, and metallicity used were 3415 K, 0.303 $R_{\odot}$, and -0.46 respectively \citep{Demangeon2021}.

We represented stellar inhomogeneities
using three parameters: fractional coverage $f_{\text{het}}$, the difference between photospheric temperature and heterogeneity temperature $\Delta T_{\text{het}}$,
and photospheric temperature $T_{\text{phot}}$. We followed the prescription used in \texttt{POSEIDON} \citep{MacDonald2017, Rathcke2021, MacDonald2023}, by using \texttt{pysynphot} \citep{pysynphot} to sample a grid of PHOENIX \citep{phoenix} models, and representing unocculted spots/faculae using stellar models with the same metallicity and log(g) values but a cooler/hotter temperature covering a fraction of the star. \footnote{\texttt{pysynphot} has a known issue with interpolating PHOENIX spectra for some stellar metallicities and temperatures. We find no evidence of the issue for the values of metallicity and temperature that we use.} The temperature
of the heterogeneity is then defined as the sum of $\Delta T_{\text{het}}$ and $T_{\text{phot}}$. This is a simplified representation of stellar inhomogeneities and does not consider time-variable stellar activity. The priors for $T_{\text{phot}}$, $\Delta T_{\text{het}}$, and $f_{\text{het}}$ are defined as $\mathcal{N}(3415, 150)$ K, $\mathcal{U}(-500, 1000)$ K, and $\mathcal{U}(0, 0.5)$ respectively.

\subsection{Constraining the Temperature}
\label{temp_constrain}
The magnitude of the features in a transmission spectrum is proportional to the vertical extent, and therefore the scale height, of the atmosphere.
The scale height is directly proportional to atmospheric temperature and 
inversely proportional to the mean molecular weight of the atmosphere -- creating a
degeneracy between temperature and mean molecular weight. 
One way to mitigate this degeneracy is to specify physically motivated prior bounds on the temperature. 
Thus, we restricted the temperature priors to within 100K of the theoretical limits 
of day-side temperature ($T_{\mathrm{d}}$) of a planet considering both extremes of circulation, 411K to 525K from \citet{Cowan2011}, giving a prior bound of 311K to 625K. 

\begin{equation}
T_{\mathrm{d}}=T_0\left(1-A_B\right)^{1 / 4}\left(\frac{2}{3}-\frac{5}{12} \varepsilon\right)^{1 / 4}
\end{equation}

where $\varepsilon$ denotes the circulation efficiency from 0 to 1, $A_{\text{B}}$ is the planetary albedo, and $T_0$ is

\begin{equation}
    T_0 = T_* \left(\frac{R_*}{a}\right)^{1/2}
\end{equation}

where $T_*$ is the stellar effective temperature, $R_*$ is the stellar radius, and $a$ is the star-planet distance. The day-side temperature is used as a conservative estimate, as the terminator temperature that we are sensitive to in transmission spectroscopy should be lower than that. 

\section{Atmospheric Retrievals} \label{sec:Atm_retr}
\subsection{Main Retrievals}
As transmission spectra of planets around M-dwarfs have been plagued by stellar contamination \citep{Lim2023, Moran2023}, for our main retrieval, we included stellar contamination. As no evidence of spot crossings was seen in the light curves, here we only modeled unocculted inhomogeneites. As as additional test, we checked whether a scenario with a planetary atmosphere but no stellar inhomogeneities can reproduce the spectrum. For this case, the priors for the parameters related to the planets atmosphere were kept the same, but the stellar parameters were removed.

We verified if a bare rock planet can still produce this spectrum if stellar inhomogeneities are present. For this scenario, we modeled the planet's contribution using only two parameters: the reference radius and the offset between NRS1 and NRS2. The priors for the stellar contribution parameters were left unchanged. We also tried a similar scenario with the planet's contribution only, essentially checking if a flat line with an offset between NRS1 and NRS2 can explain the spectrum. For each of our retrievals, we calculated a Bayesian
log-evidence value and compared it to this case. 
The values of sigma are computed using the difference in log evidence \citep{Trotta2008, Benneke2013b}.

\begin{figure*}[ht!]
\plotone{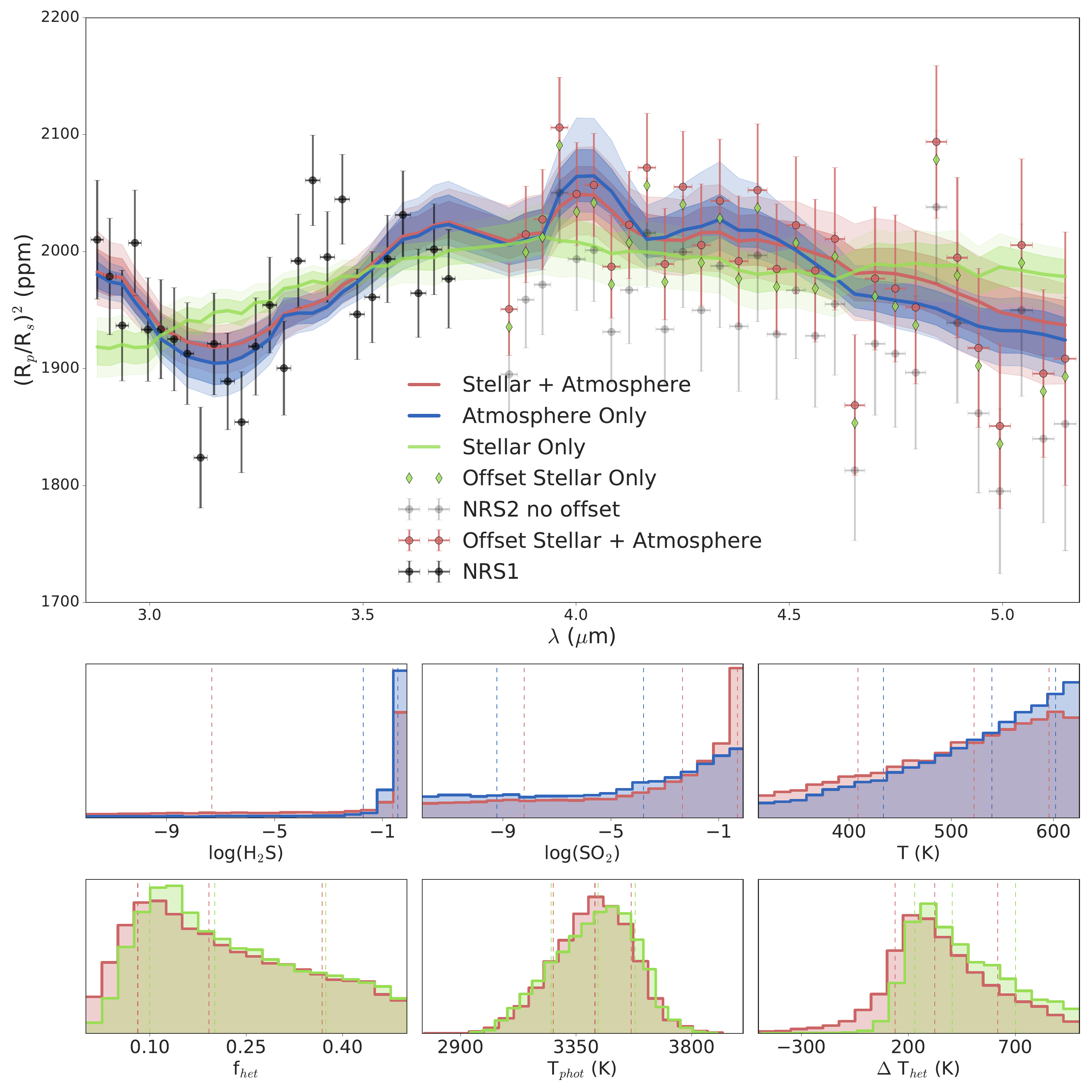}
\caption{The retrieved spectra for the main retrievals: Stellar + Atmosphere scenario is plotted in red, Atmosphere Only in blue, and Stellar Only in green. The 1$\sigma$ and 2$\sigma$ credible intervals are plotted in dark and light shades of the corresponding colors respectively. The NRS2 points are shifted by the retrieved offset, and the unshifted NRS2 points are shown in gray. Underneath the spectrum, in the first row from left to right, the retrieved posterior distributions for log(H$_2$S), log(SO$_2$), and Temperature are shown. In the second row, from left to right, the retrieved posterior distributions for $f_{\text{het}}$, $T_{\text{phot}}$, and $\Delta T_{\text{het}}$ are shown.
\label{fig:main_fig}}
\end{figure*}

\subsection{Results from Main Retrievals}
The main retrieval with stellar inhomogeneities and a planetary atmosphere favors an atmosphere with high H$_2$S and SO$_2$ abundances, with log(VMR) of $-0.62^{+0.61}_{-6.73}$ and $-2.35^{+2.05}_{-5.88}$ respectively. It also suggests the presence of unocculted stellar faculae, as $\Delta T_{\text{het}}$ is positive with a retrieved value of $324^{+295}_{-187}$K. The results from all of the retrievals are listed in Table \ref{tab:tablebig}.

The retrieval with only stellar inhomogeneities failed to provide a good fit to the observed spectrum, but resulted in similar posteriors for the stellar parameters. It is worth noting that this scenario fails to reproduce the feature at about 4 $\mu$m which can be attributed to SO$_2$. This confirms the results of Paper I that stellar activity alone cannot reproduce the observed spectrum. The scenario with the planet's contribution only, the flat line with an offset, provided a much worse fit to the spectrum - resulting in a reduced chi-squared value of only 1.86.

The fit with stellar inhomogeneities and an atmosphere was preferred at 2.24$\sigma$ to the fit with only stellar inhomogeneities, and at 3.50$\sigma$ to the fit with only a flat line and offset. The fit with only a planetary atmosphere also favors an atmosphere with high abundances of H$_2$S and SO$_2$.

The individual detection significances of H$_2$S and SO$_2$, calculated by removing each gas in turn from the main retrieval with stellar inhomogeneities and planetary atmosphere are about 2$\sigma$. Thus, these can only be interpreted as weak detections.

\subsection{Alternate Retrieval Setups}
In addition to the main retrievals, we also considered some alternate retrieval setups to test whether our choice of parameterization affects the inferences. The representation and priors for all parameters are as described in Section \ref{sec:priors}, unless otherwise stated.
 
\subsubsection{Equilibrium Chemistry Retrieval}
We performed a retrieval enforcing equilibrium chemistry using \texttt{NEMESISPY} \citep{Yang23, yangSimultaneousRetrievalOrbital2024, yangNEMESISPYPythonPackage2024}, coupled with the \texttt{FastChem} \citep{Fastchem1, FastChem}
chemical solver. We parameterized elemental abundances using three parameters: Metallicity (Z/Z$_\odot$), 
Carbon to Oxygen ratio (C/O), and Sulfur to Oxygen ratio (S/O). The prior distributions for these
parameters were set as $\mathcal{U}(0.1, 1000)$, $\mathcal{U}(0.1, 2.0)$, $\mathcal{U}(0.1, 2.0)$ respectively.
This retrieval struggled to produce a good fit to the data, and was not preferred to the baseline model of having only stellar inhomogeneities. This indicates that the atmosphere is probably not in equilibrium. The retrieved model and selected posterior distributions for some parameters are shown in Figure \ref{fig:alt_fig}.

\subsubsection{No H$_2$S/SO$_2$}\label{sec:nosulfur}
In an alternate version of the main retrieval, we removed H$_2$S and SO$_2$ from the active gases. This retrieval also struggled to produce a good fit to the data, and was not preferred to the baseline model having only stellar inhomogeneities. This result is further evidence to the tentative presence of these sulfur species in the atmosphere. The retrieved model and selected posterior distributions for some parameters are shown in Figure \ref{fig:alt_fig}.

\subsubsection{Nitrogen Background}
In another alternate version of the main retrieval, we assumed that the planet has a nitrogen dominated atmosphere as nitrogen is another gas that only has continuum features in this wavelength range. Thus, we fixed the background to N$_2$, and used log-uniform priors for the trace gases, with priors for log(VMR) set to $\mathcal{U}(-12, -0.1)$. This also resulted in significant H$_2$S and SO$_2$ abundances. This indicates that the choice of CLR priors does not bias our inferences of the atmosphere.

\subsubsection{TP Retrieval}
In this alternate version of the main retrieval, we tested the application of a non-isothermal Temperature-Pressure profile using the parameterization of \cite{madhu2009}. 
We found that this made a negligible difference to the quality of the fit, and the retrieved profile was close to isothermal, indicating that a non-isothermal profile is not statistically favored for this spectrum. This is probably as the spectrum probes a relatively narrow pressure range. This retrieval also resulted in high H$_2$S and SO$_2$ abundances.

We have also performed retrievals in which the isothermal temperature was allowed to vary between 100 K and 1000 K. In these, the retrieved temperatures were either much higher than the calculated bounds, when the atmospheric mean molecular weight was high, or much lower, for a low mean molecular weight atmosphere. 
The actual solution could possibly lie in between these extremes, and without informative prior bounds, the molecular weight and temperature are highly degenerate. 
These retrievals also recovered high H$_2$S and SO$_2$ abundances. Thus, this degeneracy does not affect the possible presence of sulfur species, which is discussed in Section \ref{sec:sulfur}. 

\begin{deluxetable*}{lcccccccc}
\tabletypesize{\scriptsize}
\tablewidth{0pt} 
\tablecaption{Priors and retrieved values of atmospheric parameters from a selection of retrievals \label{tab:tablebig}}
\tablehead{\colhead{}
            & \colhead{Priors}
            & \multicolumn{6}{c}{Posteriors}\\
            \colhead{} & 
            & \multicolumn{3}{c}{Main Retrievals}
            && \multicolumn{3}{c}{Alternate Retrievals}\\
            \cline{3-5}
            \cline{7-9}
            \colhead{} & \colhead{} &
            \colhead{Stellar + Atmosphere} & 
            \colhead{Atmosphere Only} &
            \colhead{Stellar Only} &&
            \colhead{Equilibrium} &
            \colhead{No H$_2$S/SO$_2$} &
            \colhead{FIREFLy}
            }
\startdata
log(H$_2$S) 
& CLR 
& $-0.62^{+0.61}_{-6.73}$ 
& $-0.44^{+0.42}_{-1.30}$
& -- &
& --
& --
& $-4.07^{+3.93}_{-5.11}$\\
log(SO$_2$) 
& CLR 
& $-2.35^{+2.05}_{-5.88}$ 
& $-3.79^{+2.90}_{-5.45}$
& -- &
& --
& --
& $-2.04^{+1.94}_{-6.03}$\\
log(H$_2$O) 
& CLR 
& $<-0.3$ 
& $<-2.7$
& -- &
& --
& $-5.26^{+4.33}_{-4.33}$
& $<-0.1$\\
log(CO$_2$) 
& CLR 
& $<-0.4$ 
& $<-1.8$
& -- &
& --
& $-4.88^{+4.41}_{63}$
& $<-0.1$\\
log(CO) 
& CLR 
& $<-0.3$ 
& $<-0.9$
& -- &
& --
& $-5.22^{+4.43}_{-4.38}$
& $<-0.1$\\
log(NH$_3$) 
& CLR 
& $<-1.6$ 
& $<-4.8$
& -- &
& --
& $-5.87^{+4.14}_{-4.00}$
& $<-0.4$\\
log(PH$_3$) 
& CLR 
& $<-0.4$ 
& $<-4.0$
& -- &
& --
& $-1.70^{+1.69}_{-6.27}$
& $<-0.1$\\
log(CH$_4$) 
& CLR 
& $<-1.8$ 
& $<-4.4$
& -- &
& --
& $-6.02^{+4.10}_{-3.85}$
& $<-0.9$\\
log(N$_2$) 
& CLR 
& $-5.25^{+4.43}_{-4.40}$ 
& $-5.53^{+4.38}_{-4.31}$
& -- &
& --
& $-4.76^{+4.34}_{-4.61}$
& $-5.21^{+4.40}_{-4.42}$\\
T (K)
& $\mathcal{U}(311, 625)$ 
& $522^{+73}_{-113}$ 
& $539^{+62}_{-106}$
& -- &
& $467^{+108}_{-99}$
& $458^{+105}_{-97}$
& $486^{+94}_{-110}$\\
log($P_{\text{top}}$/atm)
& $\mathcal{U}(-7, 1)$ 
& $-0.65^{+1.11}_{-2.08}$ 
& $-0.12^{+0.77}_{-0.98}$
& -- &
& $-1.28^{+1.48}_{-2.02}$
& $-2.80^{+2.41}_{-2.43}$
& $-1.73^{+1.84}_{-2.67}$\\
Power
& $\mathcal{U}(-8, 4)$ 
& $-2.28^{+4.08}_{-3.80}$ 
& $-1.78^{+3.88}_{-4.18}$
& -- &
& $-2.32^{+4.10}_{-3.61}$
& $-2.62^{+4.02}_{-3.57}$
& $-2.35^{+4.08}_{-3.67}$\\
Hazemult
& $\mathcal{U}(-5, 5)$ 
& $-2.13^{+2.75}_{-1.87}$ 
& $-2.82^{+1.66}_{-1.47}$
& -- &
& $-1.59^{+2.73}_{-2.23}$
& $0.64^{+2.96}_{-3.57}$
& $-0.98^{+3.29}_{-2.67}$\\
$f_{\text{ref}}$
& $\mathcal{U}(0.7, 1.3)$ 
& $0.97^{+0.01}_{-0.02}$ 
& $0.94^{+0.01}_{-0.01}$
& $0.99^{+0.01}_{-0.01}$ &
& $0.94^{+0.01}_{-0.01}$
& $0.97^{+0.01}_{-0.02}$
& $0.98^{+0.01}_{-0.01}$\\
Offset (ppm)
& $\mathcal{N}(0, 40)$ 
& $47.24^{+20.35}_{-18.57}$ 
& $26.16^{+16.04}_{-15.29}$
& $45.23^{+17.60}_{-18.15}$ &
& $6.86^{+15.07}_{-14.51}$
& $55.48^{+23.40}_{-22.26}$
& $15.27^{+20.34}_{-20.14}$\\
$f_{\text{het}}$
& $\mathcal{U}(0, 0.5)$ 
& $-0.19^{+0.18}_{-0.11}$ 
& --
& $0.20^{+0.17}_{-0.10}$ &
& --
& $0.24^{+0.16}_{-0.12}$
& $0.23^{+0.16}_{-0.11}$\\
$T_{\text{phot}}$
& $\mathcal{N}(3415, 150)$ 
& $3423^{+141}_{-162}$ 
& --
& $3435^{+144}_{-183}$ &
& --
& $3422^{+136}_{-177}$
& $3397^{+152}_{-166}$\\
$\Delta T_{\text{het}}$
& $\mathcal{U}(-500, 1000)$ 
& $324^{+295}_{-187}$ 
& --
& $406^{+297}_{-175}$ &
& --
& $403^{+264}_{-147}$
& $426^{+277}_{-164}$\\
Z/Z$_\odot$
& $\mathcal{U}(0.1, 1000)$ 
& -- 
& --
& -- &
& $533^{+257}_{-226}$
& --
& --\\
C/O
& $\mathcal{U}(0.1, 2.0)$ 
& --
& --
& -- &
& $0.58^{+0.49}_{-0.28}$
& --
& --\\
S/O
& $\mathcal{U}(0.1, 2.0)$ 
& -- 
& --
& -- &
& $1.16^{+0.55}_{-0.61}$
& --
& --\\
\hline
$\ln$(Z)
&  
& $462.92$ 
& $462.07$
& $461.54$ &
& $459.93$
& $461.39$
& $471.30$\\
Number of parameters
&  
& $18$ 
& $15$
& $5$ &
& $9$
& $16$
& $18$\\
MMW (amu)
&  
& $10.36$ 
& $11.07$
& -- &
& $32.14$
& $27.22$
& $30.02$\\
$\chi^2_{\text{red}}$
&  
& $1.56$ 
& $1.44$
& $1.73$ &
& $1.78$
& $1.98$
& $1.51$\\
$\sigma$
& 
& $2.24$ 
& $1.70$
& Baseline &
& Not preferred
& Not preferred
& $1.67$\\
\enddata
\tablecomments{Only 2$\sigma$ upper bounds are specified for the gas abundances that are not constrained. The detection significance $\sigma$ is computed by comparing the Bayesian Log Evidence of each model with the Stellar Only case for each reduction. For the FIREFLy retrieval, the baseline Stellar Only ln(Z) is 470.80.}
\end{deluxetable*}

\section{Discussion} \label{sec:discuss}

\subsection{Possible Presence of Sulfur Species}\label{sec:sulfur}
The primary contributors to the retrieved spectrum across all the retrieval setups we tested are invariably H$_2$S and SO$_2$, which are both prominent sulfur species.
To test their significance, we removed them from the list of active species and performed
another retrieval, otherwise identical to the main retrieval as described in Section \ref{sec:nosulfur}. This produced a much worse fit 
to the spectrum than the model with the sulfur species. By comparing the log evidences for these two scenarios, we obtain a combined detection significance of 2.32$\sigma$ for H$_2$S and SO$_2$.

Previous studies have suggested that L\,98-59\,d is a planet which is being tidally
heated \citep{seligman24}. It has a non-zero eccentricity of $0.074^{+0.057}_{-0.046}$ \citep{Demangeon2021} and the planets b, c and d have orbital resonances close to 2:4:7 \citep{seligman24}. Tidal stresses could lead to volcanic outgassing, similar to the volcanic activity on Jupiter's moon Io, and could be a possible source of H$_2$S and SO$_2$ in the atmosphere. Recent modeling studies have also suggested volcanic or outgassed origins of H$_2$S 
and SO$_2$ \citep{Claringbold2023, Tsai2024} in exoplanet atmospheres.

We find that the abundances of common atmospheric species such as H$_2$O and CO$_2$ are 
unconstrained with long tails in the posterior distribution, and we only report upper bounds 
for their abundances. The possible presence of H$_2$S also supports the absence of CO$_2$ 
and H$_2$O, as H$_2$S features in the spectrum would not be visible if 
they were present \citep{Janssen2023}. As H$_2$O is required in the photochemical production of SO$_2$, as shown in \citet{Tsai2023}, photochemistry is unlikely to be the source of SO$_2$ in the atmosphere of L\,98-59\,d. While the possible presence of H$_2$S and SO$_2$ in the absence
of H$_2$O and CO$_2$ is unlikely in equilibrium conditions \citep{Janssen2023}, they might exist
on an intensely volcanic planet.

We also do not detect the presence of any significant clouds or haze scattering. If H$_2$S and SO$_2$ are present, \citet{Jordan2021} found that these should survive above the cloud layer in the atmospheres of planets around M-dwarfs, where they could be detectable.

\subsection{Comparison between Reductions}
L\,98-59\,d has a high impact parameter \citep{Demangeon2021} and the 
transit chord crosses near to the limb of the host star from our perspective, 
thus making it difficult to pin down the limb darkening coefficients Paper I.

To check the effect of data reduction methods on our retrievals, we performed all
the same retrievals on another reduction of the spectrum using \texttt{FIREFLy}.
Here, we present a retrieval identical to our main retrieval, using this reduction instead.
While there are some differences in the retrieved parameters, the high H$_2$S
and SO$_2$ abundances are still present -- with log(VMR) of $-4.07^{+3.93}_{-5.11}$ and  $-2.04^{+1.94}_{-6.03}$ respectively. We performed a fit with a baseline stellar inhomogeneities only for this reduction as well. This 
atmospheric model is preferred at 1.67$\sigma$ to the equivalent baseline fit. The retrieved model and selected posterior distributions are shown in Figure \ref{fig:reduction_fig}.

\subsection{Comparison between Retrieval codes}
Previous studies have compared the results obtained using different retrieval codes for the same spectrum \citep{Barstow2020b, Taylor2023}. Subtle differences in the forward models used to perform retrievals can lead to significant differences in retrieved parameters. To test if our inferences are being impacted by the choice of retrieval codes, we performed a retrieval with a similar setup as our main retrieval using \texttt{Taurex3} \citep{taurex}. This retrieval, presented in Paper I, also produced similarly high abundances of H$_2$S and SO$_2$.

\begin{figure*}[ht!]
\plotone{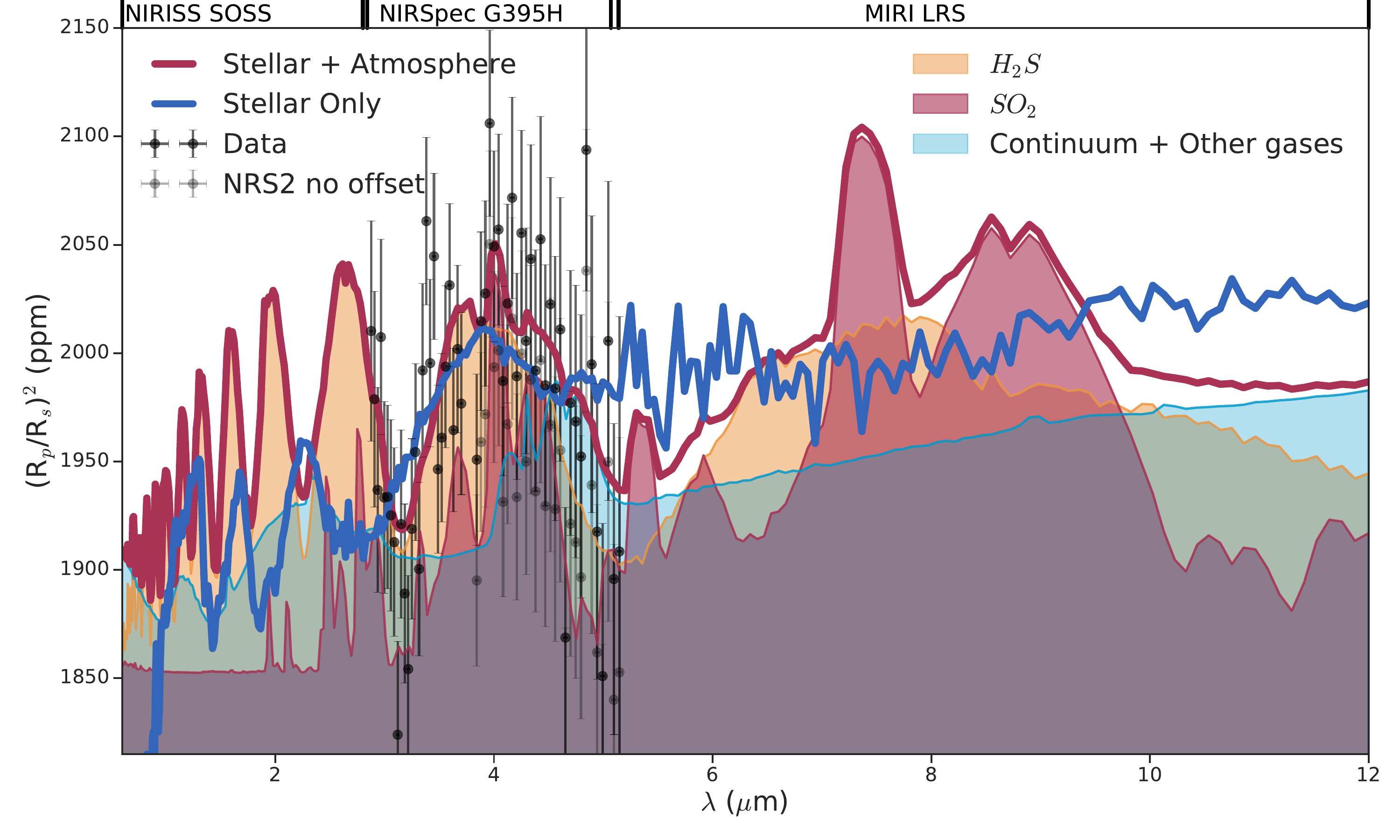}
\caption{This plot shows the retrieved spectra from the main retrieval scenarios of stellar inhomogeneities with planetary atmospheres and stellar inhomogeneities only extended to the 0.6 - 12 $\mu$m JWST wavelength range. The gas contributions for H$_2$S and SO$_2$ are highlighted. The available wavelength ranges for different JWST instruments are marked on the top of the plot. Several H$_2$S features in the NIRISS SOSS range, combined H$_2$S and SO$_2$ features in NIRSpec G395H range, and SO$_2$ features in the MIRI LRS range may be detectable in future observations.
\label{fig:extend}}
\end{figure*}

\subsection{Future Observations}
Figure \ref{fig:extend} shows the best fit model extended to between 0.6 - 12 $\mu$m. This shows H$_2$S and SO$_2$ features in other regions of the spectrum, accessible to JWST instruments NIRISS, NIRSpec and MIRI. Observations in the 0.6 - 5.0 $\mu$m range can possibly confirm or refute the signatures of these sulfur species. An accepted JWST proposal in Cycle 1: GTO 1201 (PI: Lafreniere) includes 1 transit of L\,98-59\,d with NIRISS SOSS. An accepted JWST proposal in Cycle 2: GO 4098 (PI: Benneke) also includes 1 transit of L\,98-59\,d with NIRSpec G395H and 1 transit with NIRISS SOSS. Apart from these, MIRI observations can particularly help pin down the large SO$_2$ features in the 7-10 $\mu$m range.

\section{Conclusion} \label{sec:concl}
We have presented a range of retrieval analyses for the NIRSpec/G395H spectrum of L\,98-59\,d. Our analyses favor an atmosphere with substantial amounts of sulfur species H$_2$S and SO$_2$, and an atmospheric temperature higher than the equilibrium temperature. We also find evidence of unocculted faculae on the star.
The rocky planets in the Solar System all have different atmospheric compositions, and the study of such atmospheres in exoplanetary systems could unlock a rich diversity of unexplored chemistries.

Our retrievals do not constrain the abundances of other spectrally active gases such as H$_2$O and CO$_2$, and we only report upper bounds. The bare rock with stellar inhomogeneities scenario struggles to reproduce the observed spectral features. Several H$_2$S and SO$_2$ features exist in other wavelength regions covered by JWST, as shown in Figure \ref{fig:extend}. The two scenarios of stellar inhomogeneity with an atmosphere and only stellar inhomogeneity also differ in other wavelength regions. Follow up observations in these wavelengths can confirm or refute the evidence of H$_2$S and SO$_2$, and distinguish between a planetary atmosphere and stellar inhomogeneities.

\section{Acknowledgements}
AB is supported by a PhD studentship funded by STFC and The Open University. AB thanks Dr Ryan J. MacDonald for discussions on the implementation of the CLR priors used in this work. JKB is supported by UKRI via an STFC Ernest Rutherford Fellowship (ST/T004479/1). CAH is supported by grants ST/T000295/1 and ST/X001164/1 from STFC. SRL thanks STFC for funding under grant ST/X001180/1 and UKSA under grant ST/W002949/1.
The JWST data presented in this paper were obtained from the Mikulski Archive for Space Telescopes (MAST) at the Space Telescope Science Institute. The specific observations analyzed can be accessed via  \dataset[10.17909/nrxs-cx46]{http://dx.doi.org/10.17909/nrxs-cx46}.
This work used the DiRAC Data Intensive service (DIaL2 / DIaL3) at the University of Leicester, managed by the University of Leicester Research Computing Service on behalf of the STFC DiRAC HPC Facility (www.dirac.ac.uk). The DiRAC service at Leicester was funded by BEIS, UKRI and STFC capital funding and STFC operations grants. DiRAC is part of the UKRI Digital Research Infrastructure.
We thank the anonymous referee for their detailed insights and comments which significantly improved this paper.

\bibliography{bib_aas}{}
\bibliographystyle{aasjournal}

\appendix

\section{Corner plot}
Here, we show the full corner plot for the main retrieval including stellar inhomogeneities and a planetary atmosphere. The full corner plots for all the retrievals mentioned in this paper can be found at:\url{https://github.com/riobanerjee/supplement_L9859d}.

\begin{figure*}[ht!]
\plotone{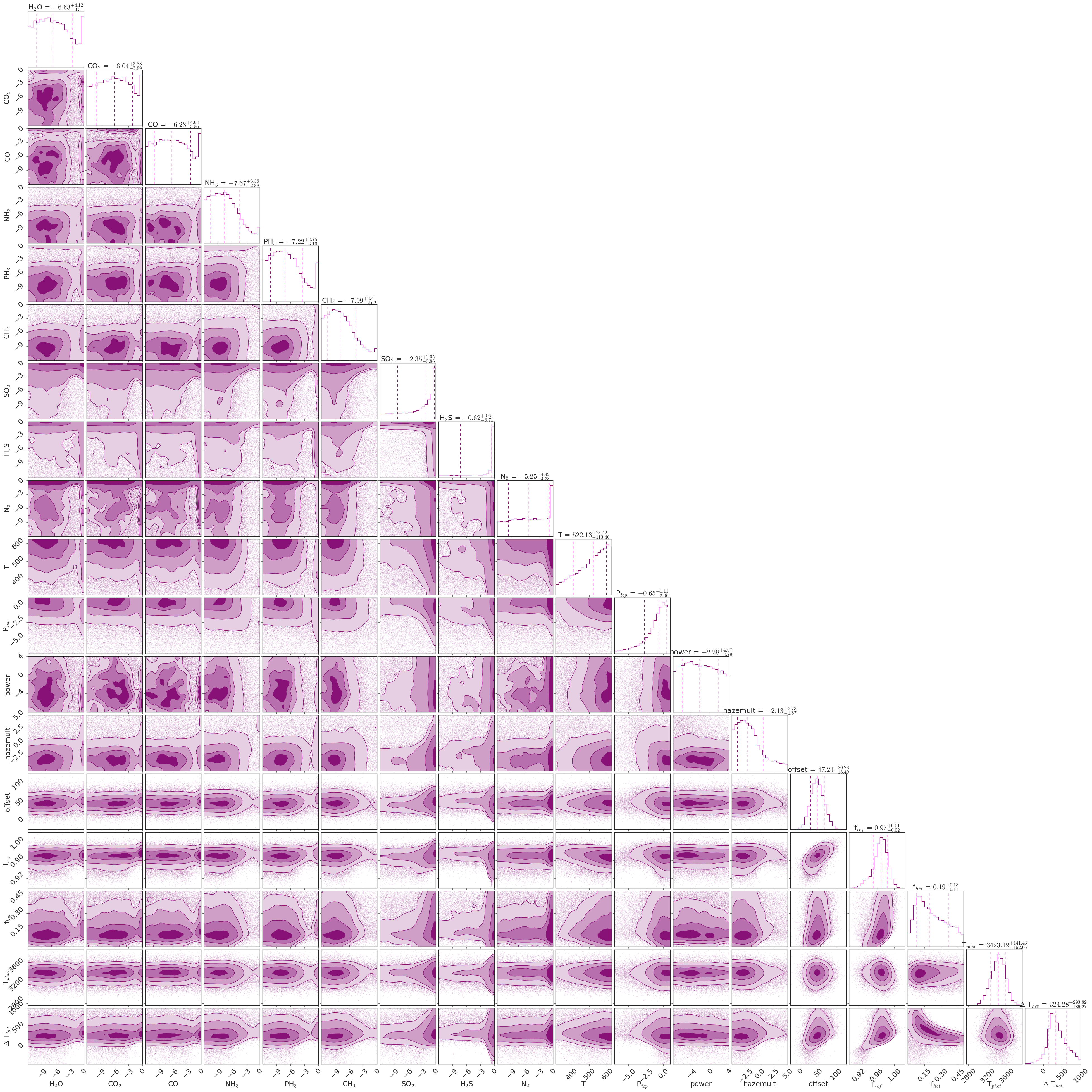}
\caption{The corner plot showing retrieved posterior distributions for each parameter for the main retrieval. The plots on the diagonal show the histograms of retrieved parameters and the inner plots show the pairwise correlations between the parameters. Large abundances for H$_2$S and SO$_2$ can be seen, with upper bounds for all other active gases.
\label{fig:corner_fig}}
\end{figure*}

\section{Alternate retrieval plot}
Here, we show the spectral fits and posteriors for selected parameters for two of the alternate retrievals: equilibrium chemistry and the retrieval without H$_2$S and SO$_2$ included.

\begin{figure}[ht!]
\plotone{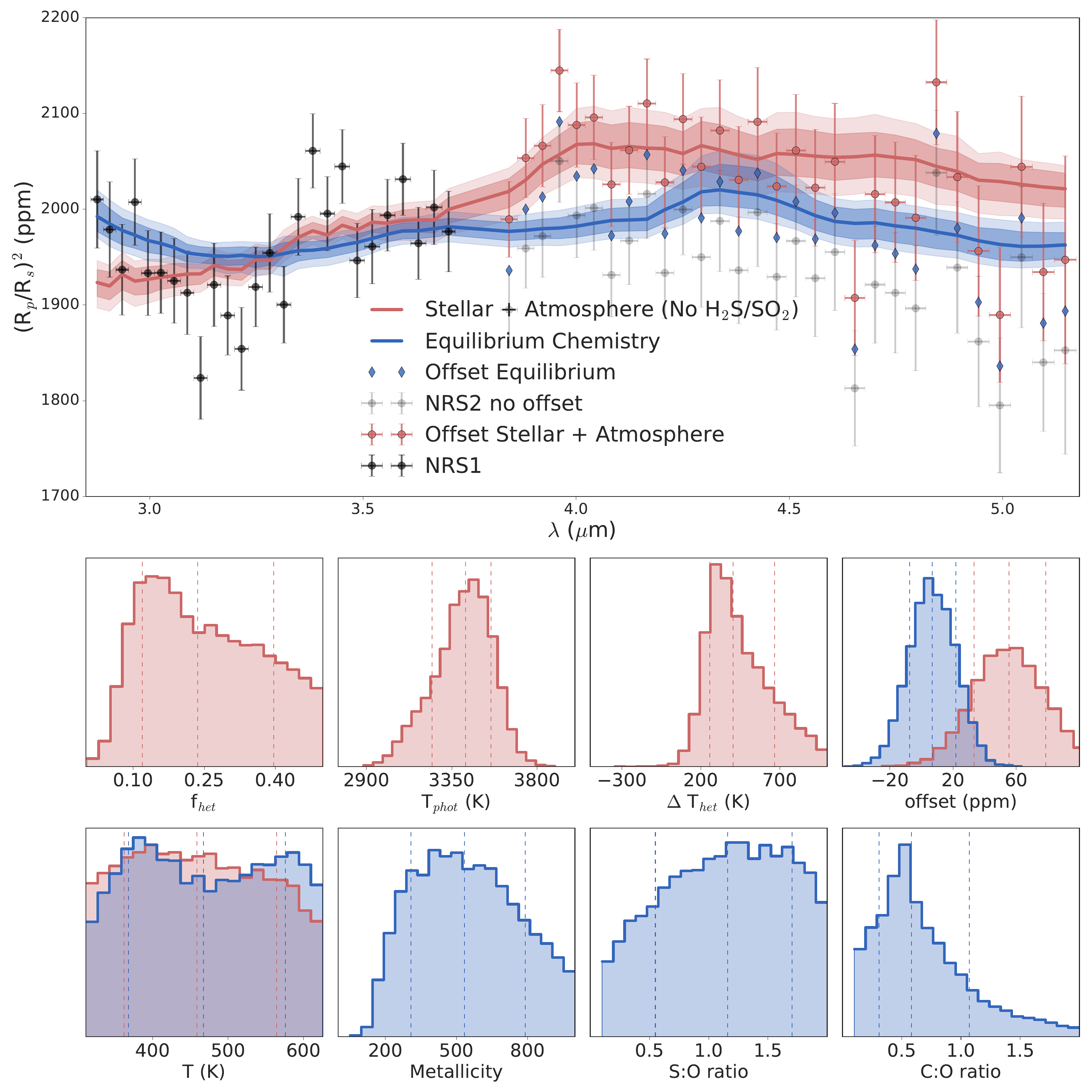}
\caption{The retrieved spectra for two the alternate retrievals are shown: The Stellar + Atmosphere scenario with H$_2$S and SO$_2$ removed is plotted in red and the equilibrium chemistry scenario is plotted in blue. The 1$\sigma$ and 2$\sigma$ credible intervals are plotted in dark and light shades of the corresponding colors respectively. The NRS2 points are shifted by the retrieved offset, and the unshifted NRS2 points are shown in gray. Underneath the spectrum, in the first row from left to right, the retrieved posterior distributions for $f_{\text{het}}$, $T_{\text{phot}}$, $\Delta T_{\text{het}}$, and offset are shown. In the second row, from left to right, the retrieved posterior distribution for Temperature, Metallicity, S:O ratio, and C:O ratio are shown.
\label{fig:alt_fig}}
\end{figure}

\section{Comparison between reductions plot}
Here, we show the spectral fits and posteriors for selected parameters for retrievals using two different reductions: \texttt{transitspectroscopy} and \texttt{FIREFLy}.
\begin{figure*}
\plotone{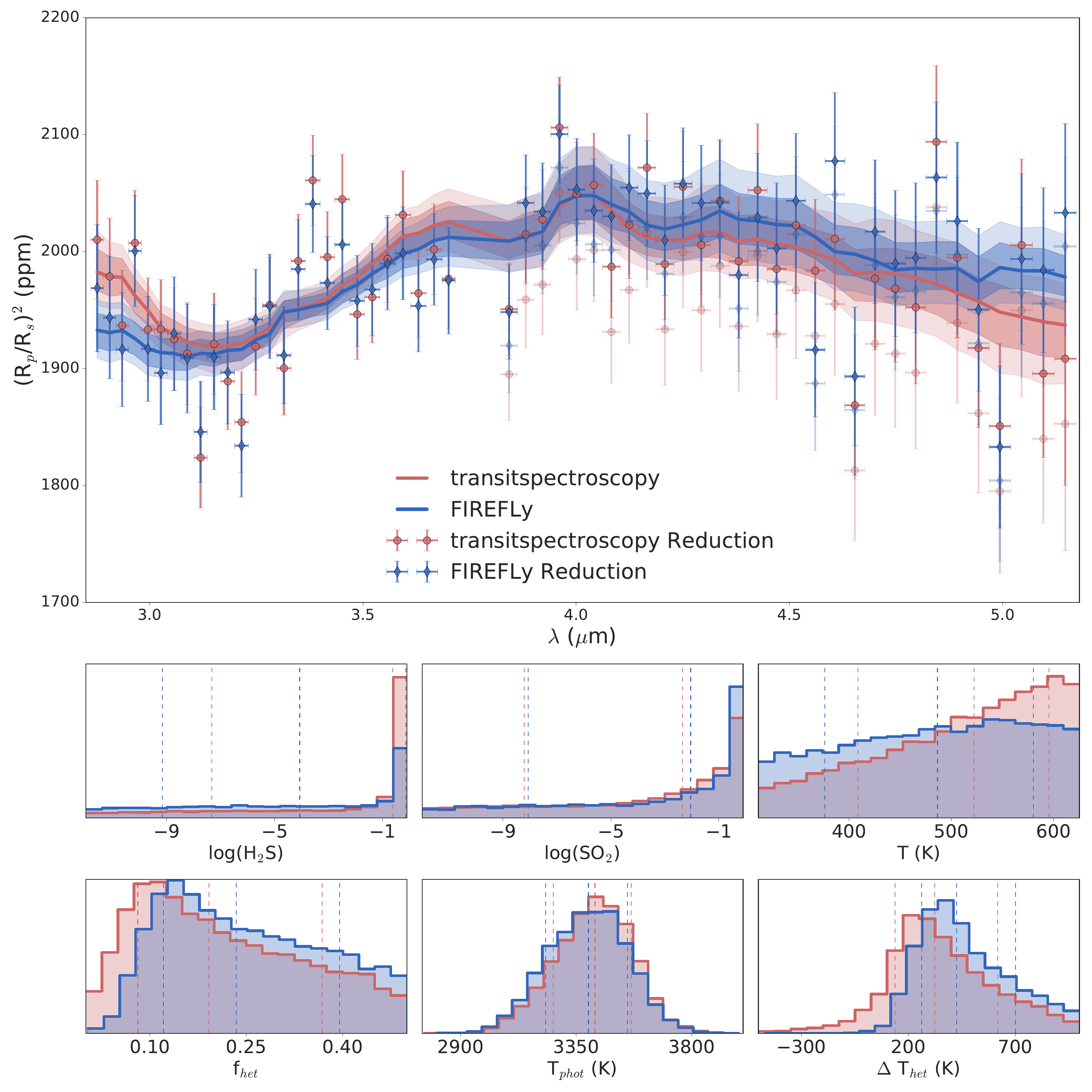}
\caption{The retrieved spectra for the retrievals using two different reductions are shown: \texttt{transitspectroscopy} is plotted in red and \texttt{FIREFLy} is plotted in blue. The 1$\sigma$ and 2$\sigma$ credible intervals are plotted in dark and light shades of the corresponding colors respectively. The NRS2 points are shifted by the retrieved offset, and the unshifted NRS2 points are shown in gray. Underneath the spectrum, in the first row from left to right, the retrieved posterior distributions for log(H$_2$S), log(SO$_2$), and Temperature are shown. In the second row, from left to right, the retrieved posterior distributions for $f_{\text{het}}$, $T_{\text{phot}}$, and $\Delta T_{\text{het}}$ are shown.
\label{fig:reduction_fig}}
\end{figure*}

\end{document}